\documentclass{article}

\usepackage{cite}
\usepackage[utf8]{inputenc}
\usepackage{fixltx2e}
\usepackage{authblk}
\usepackage{textgreek}
\usepackage{graphicx}
\usepackage{subfig}

\title{Low-power Mid-IR Supercontinuum and Rogue Wave Generation in Chalcogenide Waveguides}
\author[1,*]{S. M. Hernandez}
\author[2,3]{P. I. Fierens}
\author[1]{J. Bonetti}
\author[1,3]{D. F. Grosz}
\affil[1]{Grupo de Comunicaciones \'Opticas, Instituto Balseiro,Bariloche, R\'io Negro 8400, Argentina}
\affil[2]{Grupo de Optoelectr\'onica, Instituto Tecnol\'ogico de Buenos Aires,CABA 1106, Argentina}
\affil[3]{Consejo Nacional de Investigaciones Cient\'ificas y T\'ecnicas (CONICET),  Argentina}
\affil[*]{Corresponding author: shernandez@ib.edu.ar}

\begin{document}

\maketitle

\begin{abstract}
We present numerical results of supercontinuum (SC) generation in the mid-IR spectral region, specifically addressing the molecular fingerprint window ranging from 2.5 to 25 \textmu m. By solving the Generalized Nonlinear Schr\"odinger Equation (GNLSE) in a chalcogenide waveguide, we demonstrate low-power SC generation beyond 10 \textmu m from a pump at 5 \textmu m. Further, we investigate the short-pulse and CW regimes, and show that a simple linear dispersion profile, applicable to a broad range of chalcogenide media, is sufficient to account for the broad SC generation, and yield rich pulse dynamics leading to the frequent occurrence of rogue wave events. Results are encouraging as they point to the feasibility of producing bright and coherent light, by means of single low-power tabletop laser pumping schemes, in a spectral region that finds applications in such diverse areas as molecular spectroscopy, metrology and tomography, among others, and that is not easily addressable with other light sources.
\end{abstract}

\vspace{2pc}

\section{Introduction}

There is a need for broadband and intense light sources in the mid IR as most molecules display a distinctive spectral fingerprint in this spectral region. In particular, some spectral sub-bands within this region cannot be addressed by IR sources such as Quantum Cascade Lasers (QCLs) \cite{yao2012mid}. This problem can be tackled by supercontinuum generation through an adequate pumping scheme and choice of transmission medium. Intense and short pump pulses are best, although high CW powers can trigger SC generation as well \cite{dudley2014instabilities}. Pulsed CO\textsubscript{2} lasers are readily available and can provide intense nanosecond pulses around 10 \textmu m, and at 5 \textmu m through second-harmonic generation \cite{li2012high}. Tabletop CW CO\textsubscript{2} lasers can easily achieve kilowatt levels. CO lasers can also generate high power levels around 5 \textmu m \cite{ionin2011mode}. This leaves us with the choice of an appropriate parametric medium. Unfortunately, a well-known material such as silica (optical fibers) cannot be used, as its absorption becomes prohibitively large above 2.4 \textmu m. Chalcogenide glasses, on the other hand, display very low absorption up to 16 \textmu m \cite{shiryaev2013trends}. Other appealing characteristics of chalcogenide glasses include a zero-dispersion wavelength typically above 4-5 \textmu m (close to the pump wavelength, thus minimizing the phase mismatching of nonlinear interactions), an almost linear dispersion profile with a small dispersion slope, and a very high nonlinear index (up to 1000 times that of silica). As such, mid-IR SC generation in chalcogenide media has gathered considerable attention in recent years. There have been demonstrations of SC generation in chalcogenide step-index fibers up to 8 \textmu m by pumping at 4 \textmu m with 320-fs pulses \cite{yu2014broadband}, and from 1.4 to 13.3 \textmu m by pumping at 6.3 \textmu m with 100-fs pulses \cite{petersen2014mid}, including some impressive demonstrations in bulk glasses from 2.5 to 7.5 \textmu m, by pumping at 5.3 \textmu m with 150-fs pulses of 20 MW peak power \cite{yu2013mid}, and from 0.2 to 8 \textmu m by pumping at 1.6 \textmu m with 180-fs pulses with an astonishing 1.13 GW peak power \cite{liao2013five}. 

In this paper we focus on the study of SC generation in the mid IR, and the observation of the accompanying rogue wave phenomenon, by means of low-power pumping of a chalcogenide waveguide in both short-pulse and CW regimes. Furthermore, we show that the low and flat dispersion characteristics of chalcogenide media in the spectral region of interest, conveniently modeled by an expansion of only up to the fourth-order in the propagation constant, allow to capture the relevant pulse dynamics and to attain most of the SC spectral width in both regimes. 

\section{Simulation details}

We start by solving the Generalized Nonlinear Schr\"odinger Equation (GNLSE) expressed as \cite{dudley2010supercontinuum} 

\begin{equation}\label{eq:gnlse}
 \frac{\partial A}{\partial z} - i \sum_{n=1}^{m} \frac{i^n \beta_n}{n!} \frac{\partial^n A}{\partial t^n}
= i \left( \gamma(\omega_0) + i \gamma_1 \frac{\partial}{\partial t} \right) \left(A(z,t) \int_0^\infty R(\tau) |A(z,t-\tau)|^2 d \tau\right). 
\end{equation}

The number of terms $m$ on the left hand side is 4 in most of the simulations. Some simulations were performed using $m = 5$ in order to show that the inclusion of only four terms is enough to capture the main aspects of the resulting behavior. $\beta_2=-1$ ps\textsuperscript{2}/km, $\beta_3=0.04$ ps\textsuperscript{3}/km, $\beta_4=-0.0016$ ps\textsuperscript{4}/km, and $\beta_ 5 = 6.4\cdot 10^{-5}$ ps\textsuperscript{5}/km are the dispersion parameters, and were so chosen to be consistent with the measured dispersion profile of a chalcogenide glass in the anomalous dispersion region \cite{dantanarayana2014refractive}, i.e., a small (negative) and flat dispersion in the spectral region of interest, yielding a zero-dispersion wavelength at $\sim$4.7 \textmu m.  $\gamma(\omega_0)=100$ (W-km)$^{-1}$  and $\gamma_1=\gamma(\omega_0)/\omega_0$ are the nonlinear terms around $\omega_0$, the angular frequency corresponding to 5 \textmu m. It is important to point out that the chosen value for $\gamma$ represents a lower limit for a chalcogenide glass, consistent with a nonlinear refractive index of $\sim$100 times that of silica taking as a reference the typical effective area of a step-index silica fiber \cite{agrawal2007nonlinear,shiryaev2013trends}. The response function $R(t)$ in the convolution integral of (\ref{eq:gnlse}) is given by $(1-f_R)\delta(t) + f_R h_R(t)$, where $h_R(t)$, the Raman response of the medium, is approximated as
 
\begin{equation}\label{eq:raman}
    h_R(t) = (\tau_1^{-2} + \tau_2^{-2}) \tau_1 \exp(-t/\tau_2) \sin(t/\tau_1).
\end{equation}

\noindent $\tau_1 = 0.0155$ ps, $\tau_ 2 = 0.2305$ ps and $f_R = 0.0310$ were chosen to match the measured Raman gain spectrum, with a gain peak shift of 10 THz, in a chalcogenide glass \cite{karim2015mid}.

We consider a pump at 5 \textmu m, in two cases: short pulses with hyperbolic secant envelope $A(0,t) = \sqrt{P_0} \mbox{sech}(t/T_0)$, $T_0 = 200$ fs and $P_0 = 1$ kW; and a 1-kW continuous wave. In both cases, noise is added to the input electric field in order to achieve a realistic optical signal-to-noise ratio of 40 dB (pulsed) and 60 dB (CW), where the noise power is measured in a 50-GHz (pulsed) and a 61-GHz (CW) bandwidth. Spontaneous Raman emission is not taken into account in the analysis as its effect can be neglected compared to that of shot noise \cite{corwin2003fundamental}. Results are presented for an average of many noise realizations. Last, the waveguide is assumed to be an arbitrary 50-cm long. For a waveguide this long, attenuation can be neglected, and, thus, is not included in equation \ref{eq:gnlse},  as it can be lower than 0.6 dB/m for a chalcogenide glass fiber at the pump wavelength \cite{shiryaev2013trends}. Also, the waveguide length was chosen to guarantee stability of the results, as we verified that longer lengths and/or higher peak powers lead to numerical instability when using standard double-precision, and reasonable small step sizes, in conjunction with the RK4-IP algorithm \cite{hult2007fourth} used to solve the GNLSE.

Other simulation parameters include (resp., pulsed and CW): integration step sizes of 100 and 2.5 micra; 8192 and 16384 points for time/frequency representation; 2.4 fs and 1 fs time resolutions; and, 50 GHz and 61 GHz frequency resolutions.



\section{Results and Discussion}

\begin{figure}%
\centering
\subfloat[Pulsed (dashed line: average of 10000 realizations)]{%
\label{fig:sfig11}
\includegraphics[width=.45\linewidth]{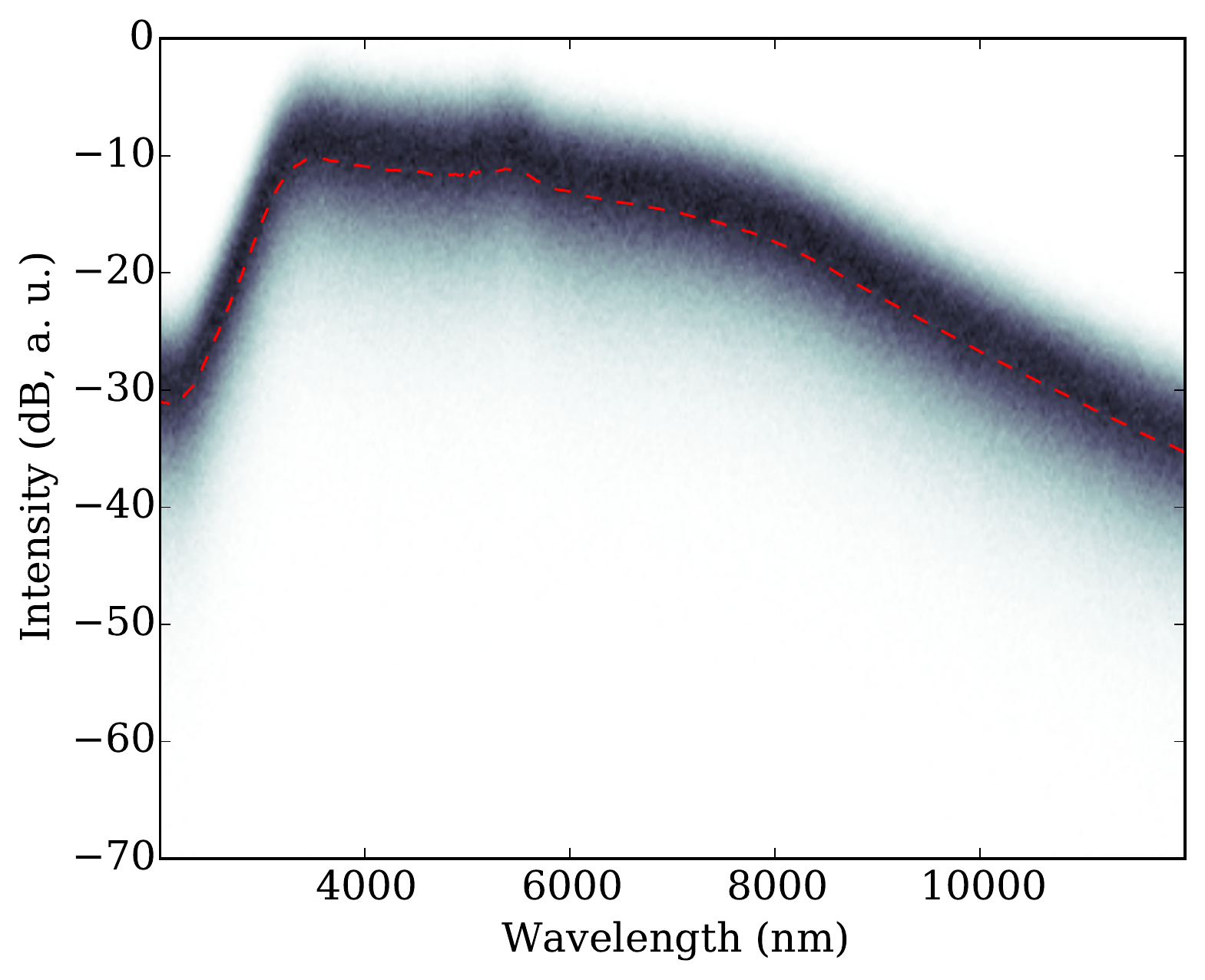}}%
\qquad
\subfloat[CW (dashed line: average of 1000 realizations)]{%
\label{fig:sfig12}
\includegraphics[width=.45\linewidth]{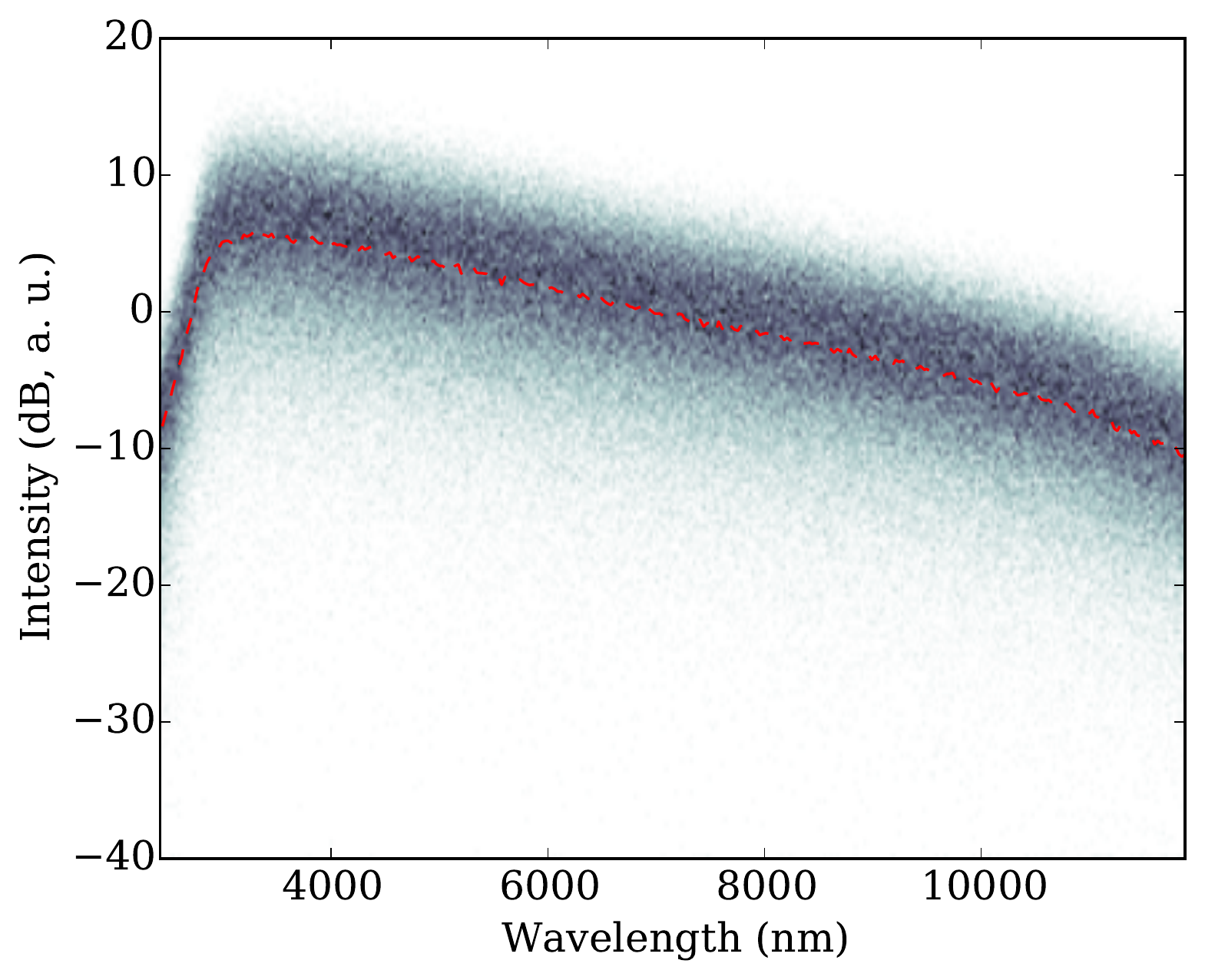}}%
\caption{Output spectra}%
\label{fig:fig1}%
\end{figure}

Output spectra are shown in Figs. \ref{fig:sfig11} and \ref{fig:sfig12} for the pulsed (average of 10000 noise realizations) and CW (average of a 1000 noise realizations) cases, respectively. We observe SC generation beyond 10 \textmu m (measured at 20-dB width). Several differences can be observed between the pulsed and CW cases.  While the maximum is above 3 \textmu m in the former case, it is close to 2 \textmu m in the latter. Moreover, in the pulsed case there is another local maximum close to 6 \textmu m which is not coincident with the pump wavelength. Finally, the spectrum decreases faster with decreasing frequency in the case of short pump pulses than in the continuous wave pump.


In order to show that the generated SC is observed at a low pump power, it is interesting to compare our findings with results in the literature. Reference \cite{yu20151} shows SC generation from 1.8 to 10 \textmu m (measured at the 40-dB point), by pumping a 11-cm chalcogenide fiber,  $\gamma= 250$ (W-km)$^{-1}$, with 330 fs pulses of 3 kW peak power produced by an Optical Parametric Amplifier (OPA). In terms of nonlinearity, the accumulated phase ($\gamma P_0 L$) is 60\% higher than in our case.  
In Ref. \cite{yu2016early}, a 10 \textmu m SC spectral width (at the 30-dB point) is demonstrated by pumping a short chalcogenide waveguide at 4.184 \textmu m with 330 fs, 4.5 kW pulses, yielding an accumulated nonlinear phase similar to that in our case. In the aforementioned references, the observed spectral width is narrower than the one reported here; we also note that, while in both papers the pump wavelength is located between two zero-dispersion wavelengths (ZDWs), we assume a simple dispersion profile with a single ZDW. Moreover, the pump wavelength and power level in these references require the use of complex OPAs, while our proposal requires ready-to-use tabletop (pulsed or CW) CO$_2$/CO lasers. Finally, simulation results in both papers use a dispersion profile modeled up to the 20th order. In this work, however, we show that it suffices to employ a simple 4th-order model capturing general dispersion characteristics of chalcogenide media. In another recent work \cite{petersen2014mid}, a comparable SC width is reported, but with a pump power at the megawatt level.

\begin{figure}%
\centering
\subfloat[Pulsed]{%
\label{fig:sfig31}
\includegraphics[width=.45\linewidth]{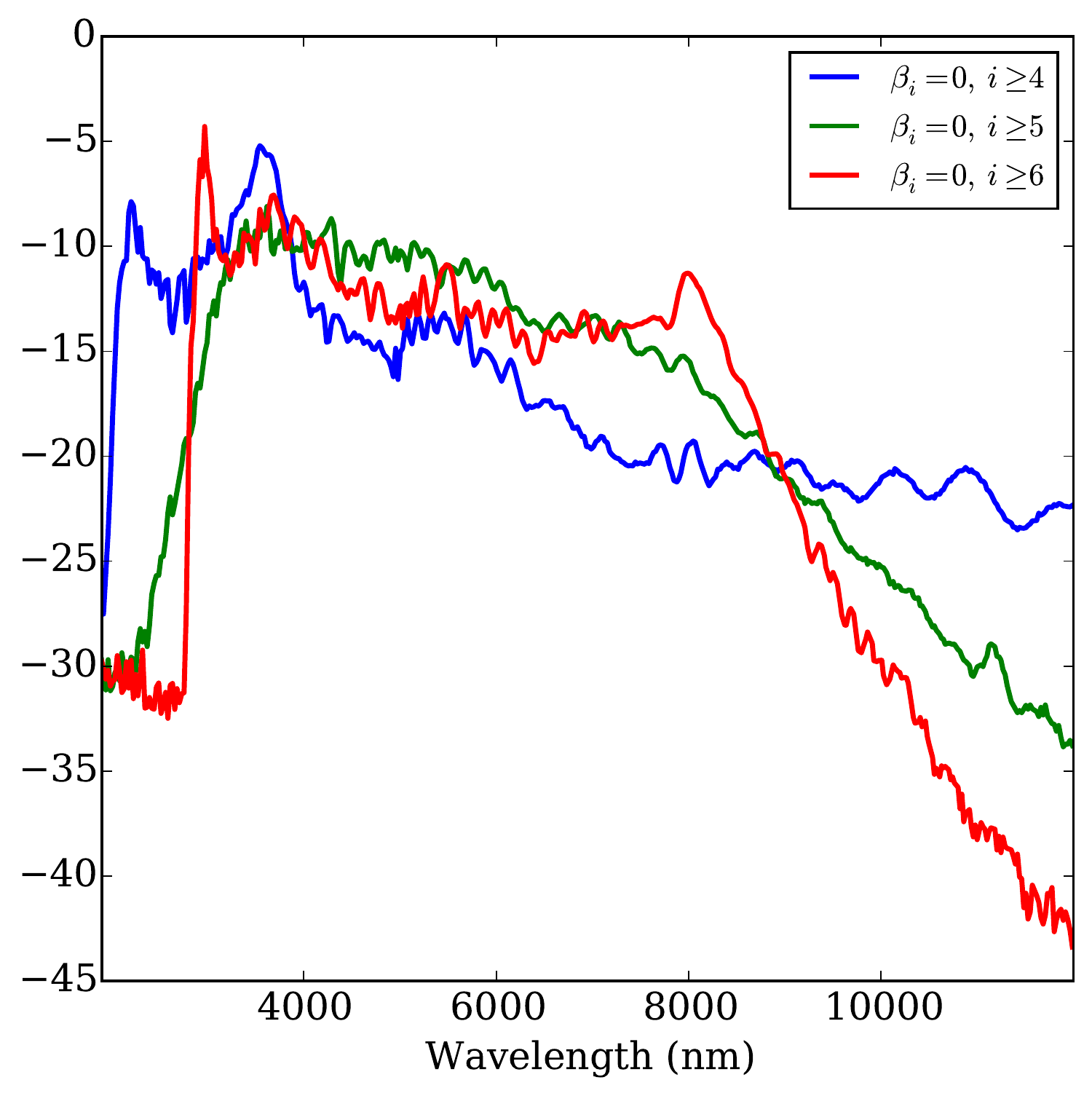}}%
\qquad
\subfloat[CW]{%
\label{fig:sfig32}
\includegraphics[width=.45\linewidth]{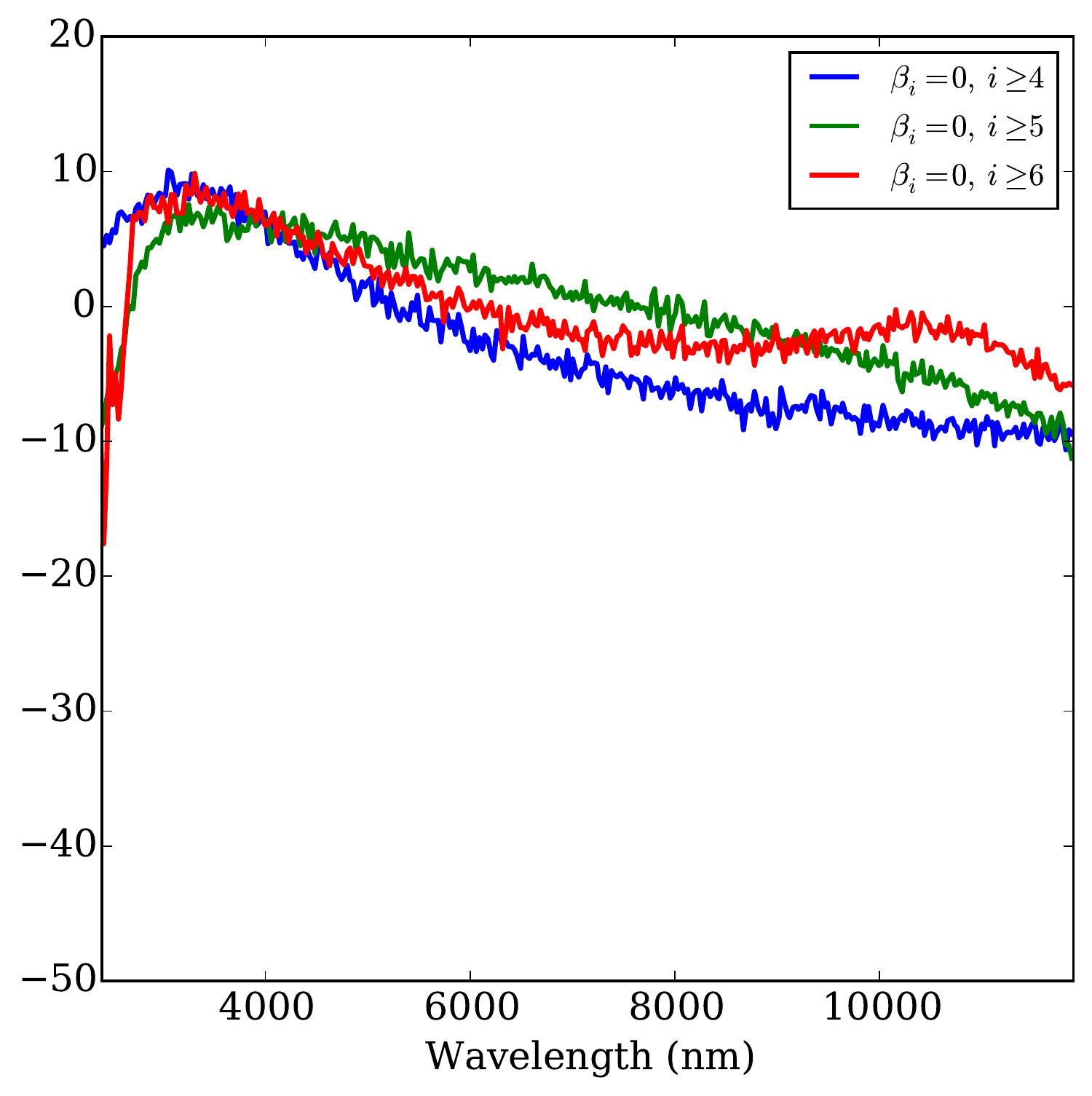}}
\caption{Effect of higher-order dispersion on the supercontinuum generation}%
\label{fig:fig3}%
\end{figure}

As mentioned before, the dispersion profile is essentially flat in the region of interest, and this leads us to expand the propagation constant up to the fourth order. To justify this choice, in Fig. \ref{fig:sfig31} we show the output spectra (pulsed case) when considering dispersion effects up to the third, fourth, and fifth order, respectively. We readily observe that the inclusion of the fifth order term leads to similar results as compared to the case where up to fourth-order dispersion is taken into account, in particular  when looking in a one-octave span. Figure \ref{fig:sfig32} shows results for the CW case, where we observe no significant differences due to higher-order dispersion.

\begin{figure}%
\centering
\subfloat[Pulsed]{%
\label{fig:sfig21}
\includegraphics[width=.45\linewidth]{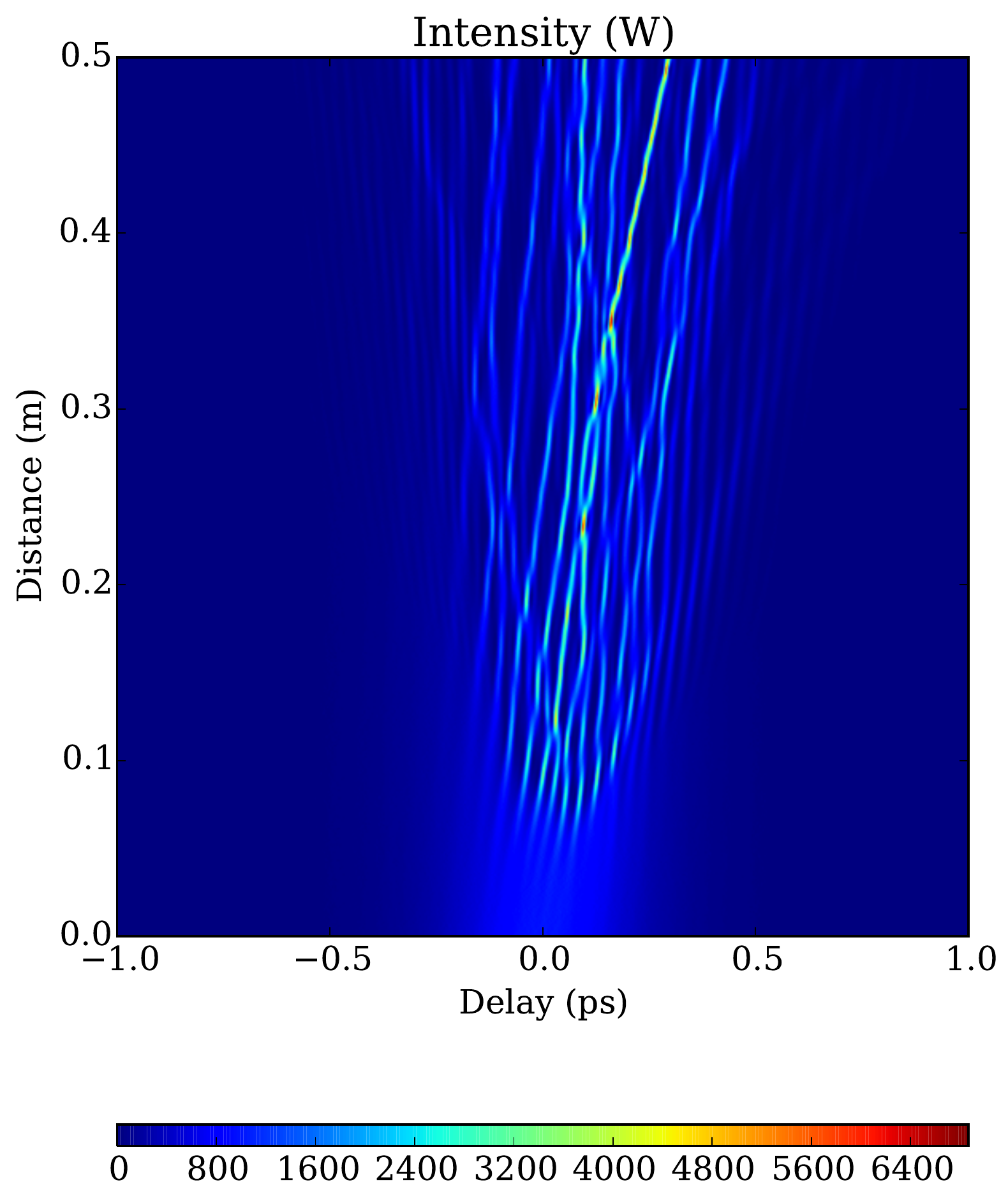}}%
\qquad
\subfloat[CW]{%
\label{fig:sfig22}
\includegraphics[width=.45\linewidth]{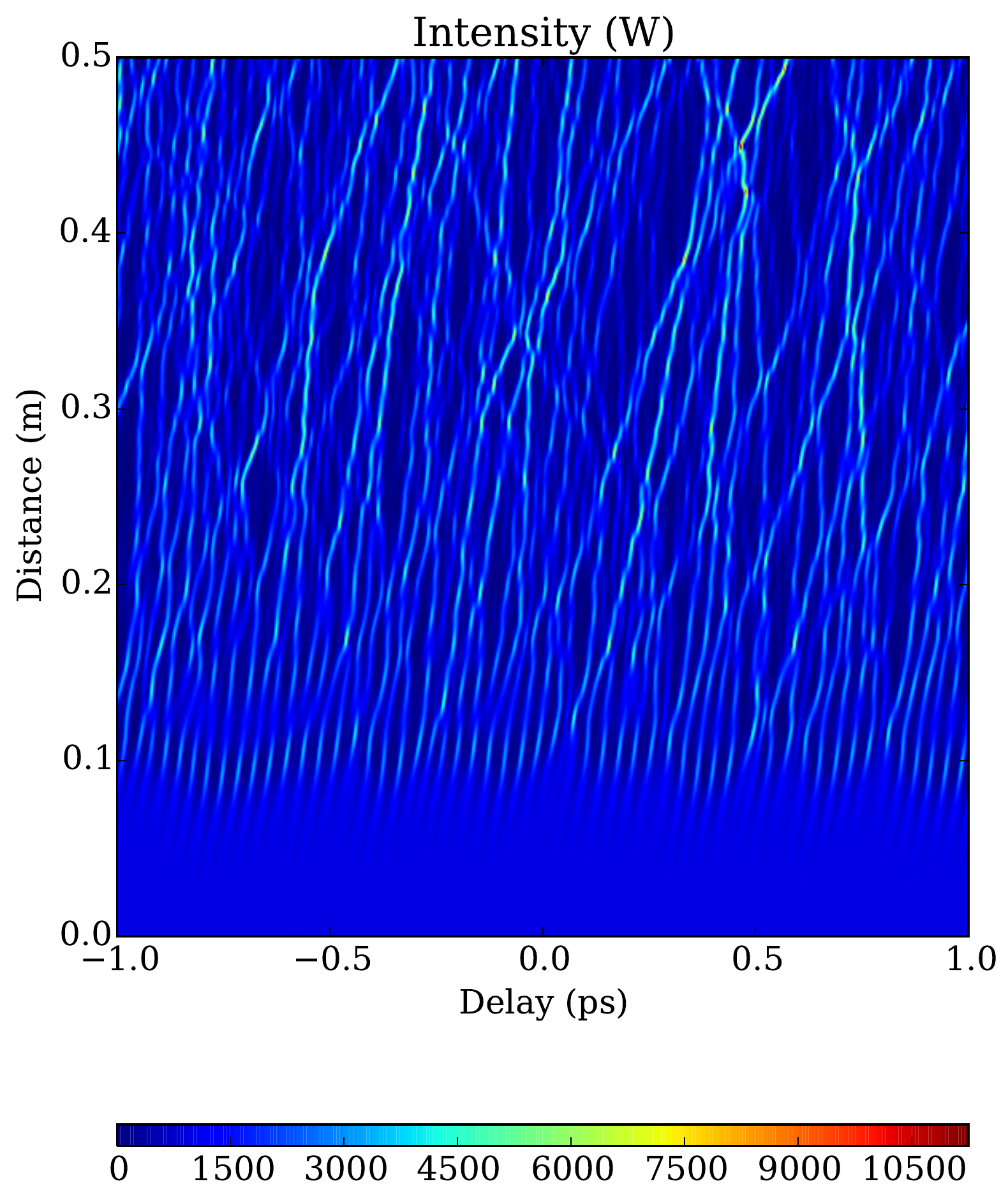}}
\caption{Time evolution along the waveguide}%
\label{fig:fig2}%
\end{figure}

Figure \ref{fig:fig2} shows the time evolution along the waveguide, displaying the complex pulse interaction expected from soliton dynamics (Fig. \ref{fig:sfig21}, pulsed case) and the signature of pulse trains and ensuing soliton dynamics characteristic of modulation instability-induced SC generation (Fig. \ref{fig:sfig22}, CW case). It is interesting to note that the emergence of new pulses occurs at approximately the same traveled distance into the fiber, $\sim 8$ cm, in both cases.

\begin{figure}%
\centering
\subfloat[Evolution of the spectrum]{%
\label{fig:sfig51}
\includegraphics[width=.45\linewidth]{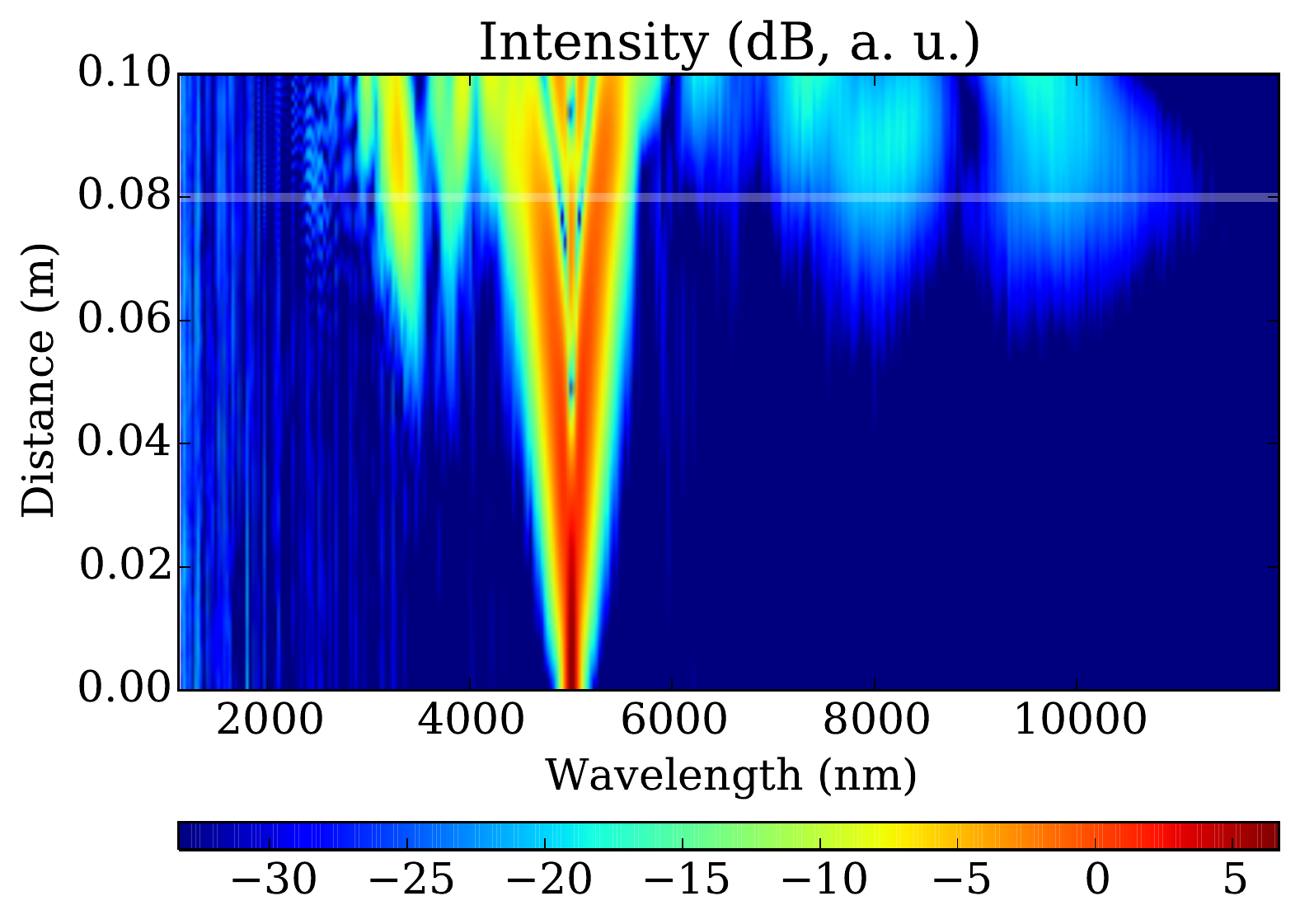}}%
\qquad
\subfloat[Spectrum at 8 cm]{%
\label{fig:sfig52}
\includegraphics[width=.45\linewidth]{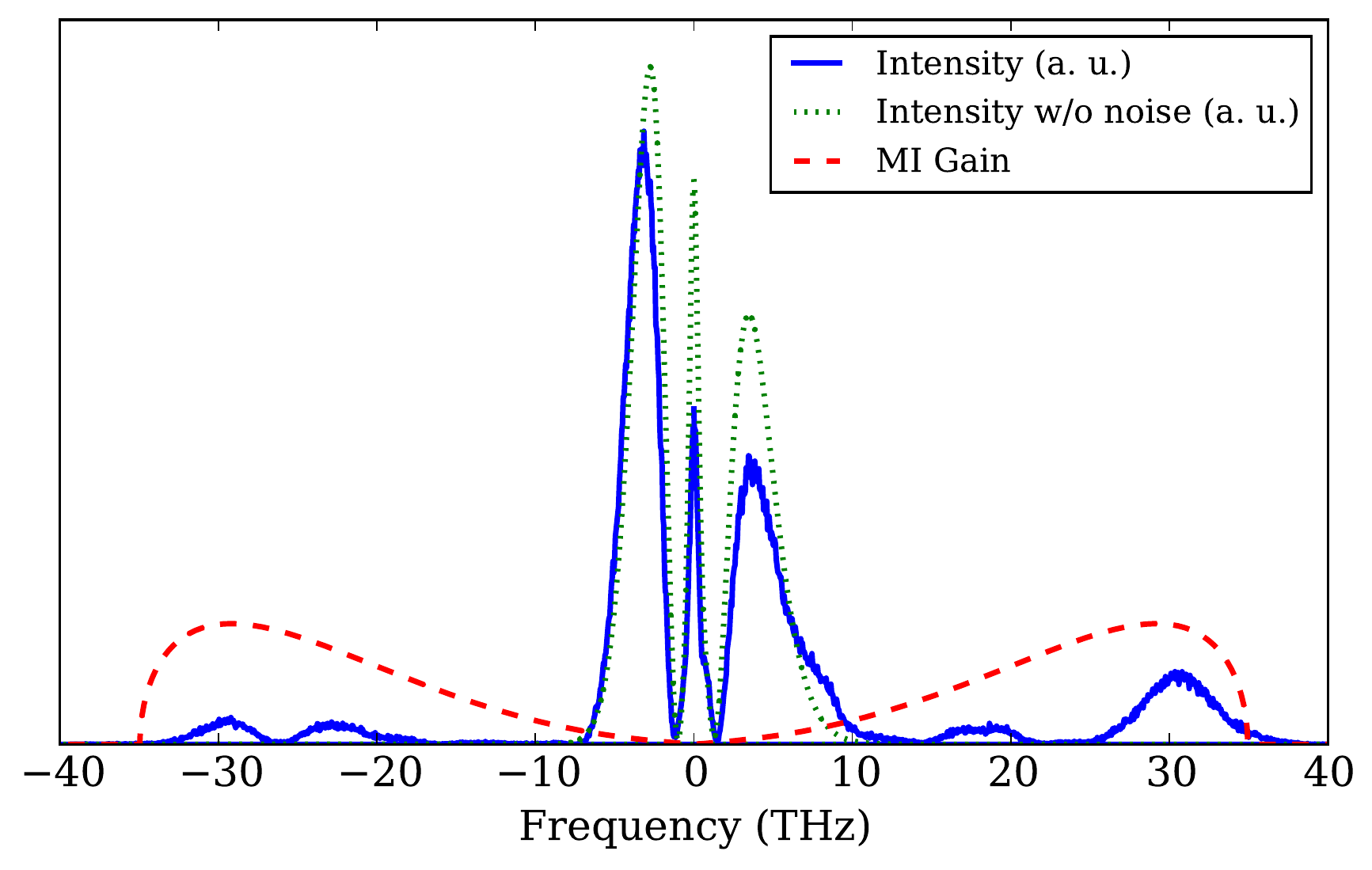}}
\caption{Spectrum in the pulsed-pump case. (a) Evolution along the waveguide. (b) Power spectrum at 8 cm in the noisy-input case (solid) and the noiseless case (dotted), and modulation instability gain (dashed).}%
\label{fig:fig5}%
\end{figure}

Although one may be tempted to propose a soliton fission mechanism for the initial formation of sub-pulses in the pulsed-pump case, a few calculations show that this cannot be the dominant factor. The input pulse corresponds to a soliton of order $N = \sqrt{L_D/L_{NL}} \sim 63$ \cite{agrawal2007nonlinear}. The soliton fission length is usually estimated as $L_{\mathrm{fiss}}\sim L_D/N \sim 63$ cm \cite{dudley2010supercontinuum}, which is longer than the waveguide. Thus, modulation instability appears to be the cause of the initial optical pulse breakdown. Fig. \ref{fig:sfig51} shows the evolution of the spectrum along the waveguide for the pulsed case. The symmetrical broadening in the form of a `V' shape, observed in the first few centimeters into the waveguide, is a typical signature of self-phase modulation (SPM). Fig. \ref{fig:sfig52} shows the average spectrum at 8 cm together with the spectrum in the case where the input pump is noiseless. It also shows the gain spectrum due to modulation instability (MI),  calculated taking into account high-order dispersion \cite{sorensen2012seededsc}, as a direct calculation using the standard formula ($\omega_{\mathrm{max}} = \sqrt{2\gamma P_0/|\beta_2|}$, see \cite{agrawal2007nonlinear}) yields a higher frequency offset. Taking into account that the sub-pulses that emerge near the input end of the waveguide are approximately 30 fs wide (Fig. \ref{fig:sfig22}) and noting that non-negligible spectral content at $\pm$ 30 THz appears only in the noisy-input case (Fig. \ref{fig:sfig52}), it can be concluded that the initial breakdown into pulses is due to noise-seeded modulation instability.

Similarly, in the CW case, noise-seeded modulation instability leads to the formation of a pulse train with a period $\sim$ 30 fs (Fig. \ref{fig:sfig22}) consistent with the location of the maxima of the MI-gain spectrum (Fig. \ref{fig:sfig52}).

\begin{figure}%
\centering
\subfloat[Pulsed]{%
\label{fig:sfig41}
\includegraphics[width=.45\linewidth]{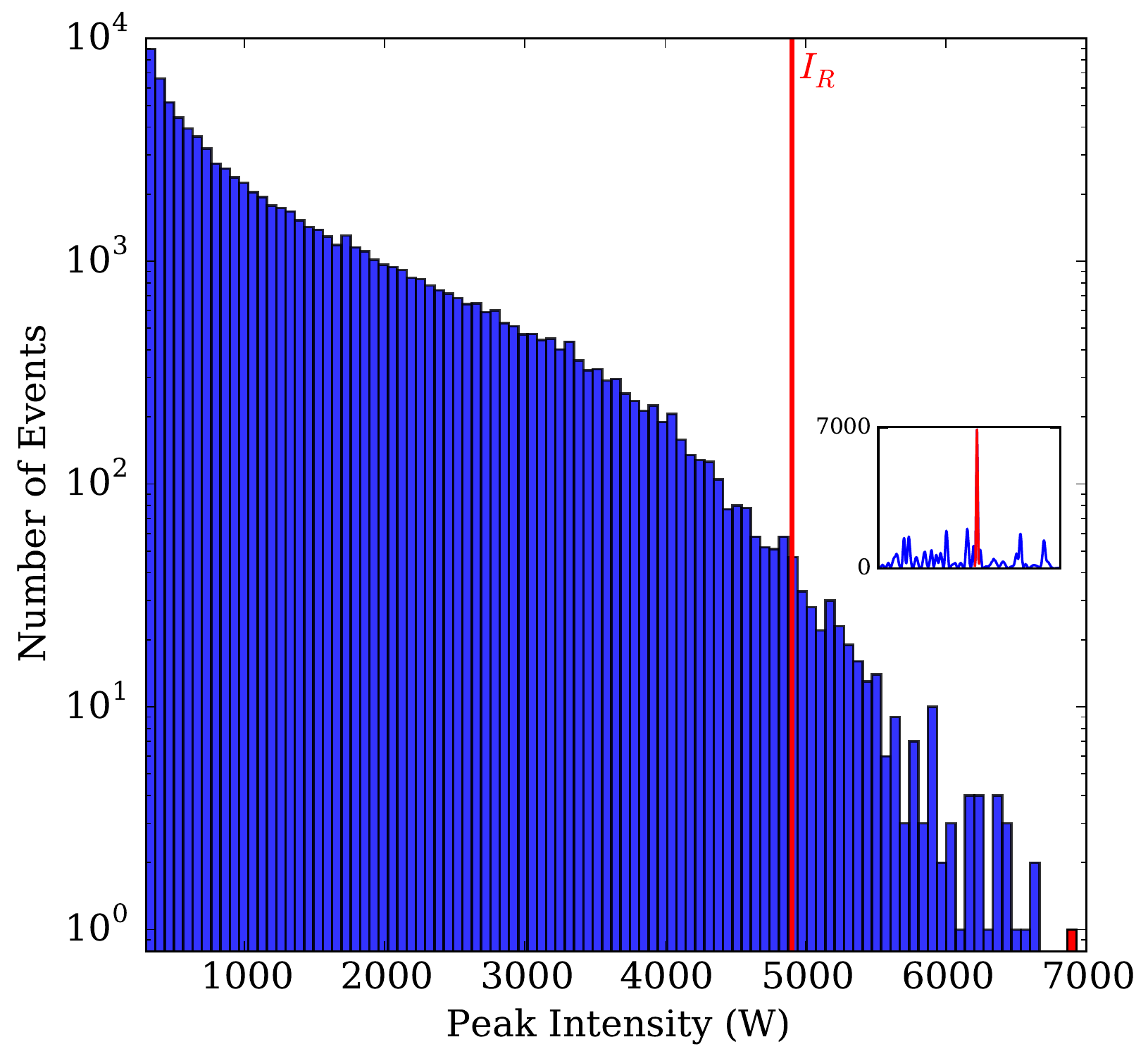}}%
\qquad
\subfloat[CW]{%
\label{fig:sfig42}
\includegraphics[width=.45\linewidth]{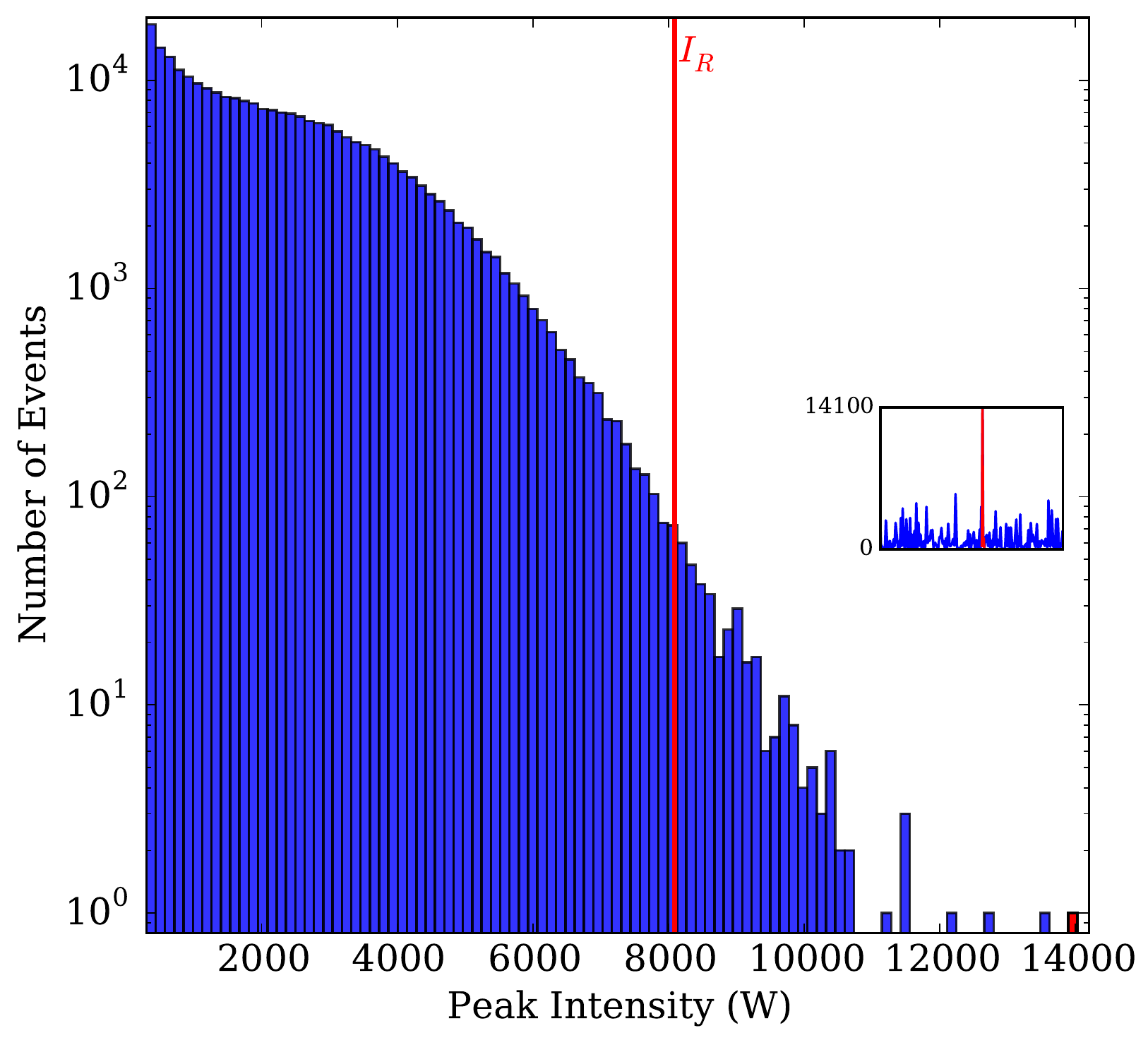}}
\caption{Intensity histogram at the waveguide output. The vertical line marks the threshold for rogue events. Inset: detail of the highest-intensity event.}%
\label{fig:fig4}%
\end{figure}

From the previous results, and the large amount of noise realizations, it is only natural to explore the possible occurrence of rogue events. Figures \ref{fig:sfig41} and \ref{fig:sfig42} show results obtained for pulse-intensity statistics for 5000 (pulsed) and 500 (CW) noise realizations, respectively. It is worth noting that the CW case produces a much larger number of pulses for each realization as compared to the pulsed case; as such, a significant number of events can be attained with a relatively small number of noise realizations. In both cases, the classical signature of a long tail ('L shape') distribution is observed, pointing at the enhanced probability of extreme events. Specifically, by choosing the usually adopted threshold for such events \cite{solli2007optical}, namely the so-called 'significant wave height', corresponding to twice the average of the upper third of the intensity distribution, we find that approximately 0.3\% and 0.15\% correspond to rogue waves for the pulsed and CW cases, respectively. Insets in Figures  \ref{fig:sfig41} and \ref{fig:sfig42} show examples of some of the highest-power resulting pulses, $\sim$ 7 kW in the pulsed case and $\sim$ 14 kW in the continuous wave case. It is interesting to emphasize that we observe the occurrence of these extreme events resorting to a very simple dispersion profile, as compared to results in the literature that include up to 10th-orden dispersion, and at similar levels of nonlinearity \cite{dudley2008harnessing}.

\section{Conclusions}
We showed, through numerical modeling, the feasibility of supercontinuum generation in the mid~IR by pumping a chalcogenide waveguide with a pump wavelength at 5 \textmu m, both in the short pulse and CW regimes. Furthermore, in both cases initial breakdown into short pulses appears to be due to noise-seeded modulation instability, and supercontinuum generation proceeds through soliton dynamics and Raman self-frequency shift.

We found substantial spectral content beyond 10 \textmu m, a promising result as it points to the feasibility of covering a larger part of the mid-IR window by means of SC generation through low-power pumping readily available from CO\textsubscript{2}/CO lasers.

Further, we investigated the role of the dispersion profile of the chalcogenide medium, showing that inclusion of only up to the fourth-order term is sufficient to account for the observed rich pulse dynamics and SC spectral width within one-octave span. 

Extensive simulation of noise realizations in both pulsed and CW cases allowed us to study some of the statistical properties of the output pulses, and found evidence of rogue wave events. Deeper analysis and control of such events may lead to more stable SC sources for real mid-IR applications such as molecular spectroscopy, metrology and tomography, among others.

\section*{Acknowledgements}
The authors wish to thank E. Cortizo, I. Tarnapolsky, and J. Ben\'{i}tez for enlightening discussions.


\begin{thebibliography}{10}
\expandafter\ifx\csname url\endcsname\relax
  \def\url#1{{\tt #1}}\fi
\expandafter\ifx\csname urlprefix\endcsname\relax\def\urlprefix{URL }\fi
\providecommand{\eprint}[2][]{\url{#2}}

\bibitem{yao2012mid}
Yao Y, Hoffman A~J and Gmachl C~F 2012 {\em Nature Photonics\/} {\bf 6}
  432--439

\bibitem{dudley2014instabilities}
Dudley J~M, Dias F, Erkintalo M and Genty G 2014 {\em Nature Photonics\/} {\bf
  8} 755--764

\bibitem{li2012high}
Li D, Guo J, Yang G, Meng F, Zhang L, Xie J, Chen F, Shao C, Zhang C, Geng Y
  {\em et~al.\/} 2012 {\em Laser Physics\/} {\bf 22} 725--729

\bibitem{ionin2011mode}
Ionin A, Guo J, Zhang L~M, Xie J~J, Andreev Y~M, Kinyaevsky I, Klimachev Y~M,
  Kozlov A~Y, Kotkov A, Lanskii G {\em et~al.\/} 2011 {\em Laser Physics
  Letters\/} {\bf 8} 723--728

\bibitem{shiryaev2013trends}
Shiryaev V and Churbanov M 2013 {\em Journal of Non-Crystalline Solids\/} {\bf
  377} 225--230

\bibitem{yu2014broadband}
Yu Y, Gai X, Ma P, Choi D~Y, Yang Z, Wang R, Debbarma S, Madden S~J and
  Luther-Davies B 2014 {\em Laser \& Photonics Reviews\/} {\bf 8} 792--798

\bibitem{petersen2014mid}
Petersen C~R, M{\o}ller U, Kubat I, Zhou B, Dupont S, Ramsay J, Benson T,
  Sujecki S, Abdel-Moneim N, Tang Z {\em et~al.\/} 2014 {\em Nature
  Photonics\/} {\bf 8} 830--834

\bibitem{yu2013mid}
Yu Y, Gai X, Wang T, Ma P, Wang R, Yang Z, Choi D~Y, Madden S and Luther-Davies
  B 2013 {\em Optical Materials Express\/} {\bf 3} 1075--1086

\bibitem{liao2013five}
Liao M, Gao W, Cheng T, Xue X, Duan Z, Deng D, Kawashima H, Suzuki T and Ohishi
  Y 2013 {\em Applied Physics Express\/} {\bf 6} 032503

\bibitem{dudley2010supercontinuum}
Dudley J~M and Taylor J~R 2010 {\em Supercontinuum generation in optical
  fibers\/} (Cambridge University Press)

\bibitem{dantanarayana2014refractive}
Dantanarayana H~G, Abdel-Moneim N, Tang Z, Sojka L, Sujecki S, Furniss D,
  Seddon A~B, Kubat I, Bang O and Benson T~M 2014 {\em Optical Materials
  Express\/} {\bf 4} 1444--1455

\bibitem{agrawal2007nonlinear}
Agrawal G~P 2012 {\em Nonlinear fiber optics\/} 5th ed (Academic press)

\bibitem{karim2015mid}
Karim M, Rahman B and Agrawal G~P 2015 {\em Optics express\/} {\bf 23}
  6903--6914

\bibitem{corwin2003fundamental}
Corwin K, Newbury N~R, Dudley J, Coen S, Diddams S, Washburn B, Weber K and
  Windeler R 2003 {\em Applied Physics B\/} {\bf 77} 269--277

\bibitem{hult2007fourth}
Hult J 2007 {\em Journal of Lightwave Technology\/} {\bf 25} 3770--3775

\bibitem{yu20151}
Yu Y, Zhang B, Gai X, Zhai C, Qi S, Guo W, Yang Z, Wang R, Choi D~Y, Madden S
  {\em et~al.\/} 2015 {\em Optics letters\/} {\bf 40} 1081--1084

\bibitem{yu2016early}
Yu Y, Gai X, Ma P, Vu K, Yang Z, Wang R, Choi D~Y, Madden S and Luther-Davies B
  2016 {\em to appear in Optics letters\/}

\bibitem{sorensen2012seededsc}
S{\o}rensen S~T, Larsen C, M{\o}ller U, Moselund P~M, Thomsen C~L and Bang O
  2012 {\em JOSA B\/} {\bf 29} 2875--2885

\bibitem{solli2007optical}
Solli D, Ropers C, Koonath P and Jalali B 2007 {\em Nature\/} {\bf 450}
  1054--1057

\bibitem{dudley2008harnessing}
Dudley J~M, Genty G and Eggleton B~J 2008 {\em Optics Express\/} {\bf 16}
  3644--3651

\end{thebibliography}

\providecommand{\newblock}{}

\end{document}